\def\Missing#1#2{
         \ifmmode
              {#1}\kern-0.6em\lower-.1ex\hbox{/}_{#2} 
         \else
             ${#1}\kern-0.6em\lower-.1ex\hbox{/}_{#2}$
         \fi}
\def\met{\mbox{${\hbox{$E$\kern-0.6em\lower-.1ex\hbox{/}}}_T$}} 
\def\D0{D\O}                            
\def\d0draft{}
\def\err#1#2#3 {{\it Erratum} {\bf#1},{\ #2} (19#3)}
\def\ib#1#2#3 {{\it ibid.} {\bf#1},{\ #2} (19#3)}
\def\nc#1#2#3 {Nuovo Cim. {\bf#1} ,#2(19#3)}
\def\nim#1#2#3 {Nucl. Instr. Meth. {\bf#1},{\ #2} (19#3)}
\def\np#1#2#3 {Nucl. Phys. {\bf#1},{\ #2} (19#3)}
\def\pl#1#2#3 {Phys. Lett. {\bf#1},{\ #2} (19#3)}
\def\prev#1#2#3 {Phys. Rev. {\bf#1},{\ #2} (19#3)}
\def\prl#1#2#3 {Phys. Rev. Lett. {\bf#1},{\ #2} (19#3)}
\def\rmp#1#2#3 {Rev. Mod. Phys. {\bf#1},{\ #2} (19#3)}
\def\zp#1#2#3 {Zeit. Phys. {\bf#1},{\ #2} (19#3)}

\input{epsf}
\documentclass[epj]{svjour}
\usepackage[spanish]{babel}  
\usepackage[latin1]{inputenc}  
\usepackage{isolatin1}   
\usepackage{graphics}
\usepackage{amssymb,amsfonts, amsbsy}
\usepackage[T1]{fontenc}
\usepackage[latin1]{inputenc}

\begin{document}

\title{Una prueba empírica de generadores de números pseudoaleatorios mediante un proceso de decaimiento exponencial.}
\author{H.F. Coronel-Brizio\inst{1}\  \thanks{\emph{e-mail:} hcoronel@uv.mx}, 
A.R. Hernández-Montoya\inst{1}\ \thanks{\emph{e-mail:} alhernandez@uv.mx}
M.A. Jiménez-Montaño\inst{1}\  \thanks{\emph{e-mail:} majm@uv.mx} 
y L.E. Mora Forsbach\inst{2}\  \thanks{\emph{e-mail:} emora@cimat.mx}
}%

\institute{Facultad de Física e Inteligencia Artificial.
Universidad Veracruzana, Apdo. Postal 475. Xalapa, Veracruz. México. \and Centro de Investigación en Matemáticas.
CIMAT. Apdo.Postal  402, Guanajuato, Gto., C.P. 36000, México}

\date{Received: date / Revised version: date}
\abstract{
\noindent
Las pruebas empíricas que usan procesos o modelos físicos para probar generadores de números
 pseudoaleatorios, complementan las pruebas de aleatoriedad teóricas y han sido usadas con mucho éxito.
En este trabajo, se presenta una metodología estadística para evaluar la calidad
de generadores de números pseudoaleatorios, ilustrando
el método en el contexto del proceso de decaimiento radiactivo y utilizando para ello
algunos generadores de uso común  en la Física.
\\
\\
Empirical tests for pseudorandom number generators based on the use of processes or physical models
have been successfuly used and are considered as complementary to theoretical test of randomness.
In this work a statistical methodology for evaluating the quality of pseudorandom number generators
is presented. The method is illustrated in the context of the so-called exponential decay process,
using some pseudorandom number generators commonly used in physics.
} 
\authorrunning{ Coronel-Brizio, Hernandez-Montoya }%
\titlerunning{Una prueba empírica de generadores de números pseudoaleatorios}

\maketitle
\section{Introducción}
\label{intro}
\vspace*{-.4cm}

\noindent
La simulación  mediante el Método de Monte Carlo  \cite{MC1,MC2} hace uso intensivo de sucesiones
 de números aleatorios y  es una técnica estándar ampliamente aplicada desde hace mas de 50 años en diversas
 ramas de la Física, especialmente en la Física de altas energías y en la Física Estadística.
 Además de su aplicabilidad en el  método de Monte Carlo, el  uso y obtención de sucesiones de números aleatorios
  constituye hoy en día una investigación muy activa con aplicaciones en campos tan diversos  como la criptografía, integración montecarlo, ecología, identificación biométrica y aún Inteligencia Artificial  \cite{stat1,crypto1,integracion,eco1,identificai,identifica2,inteligencia}. \\
\noindent
Estrictamente hablando, obtener sucesiones de números realmente aleatorios implica la utilización
de algún fenómeno físico de naturaleza estocástica como el arrojar una moneda al aire,
el ruido de un circuito electrónico, el decaimiento de un material radioactivo, el conteo de fotones mediante detectores 
centelladores y mas recientemente, se han propuesto métodos menos tradicionales basados en fenómenos tales como el flujo turbulento de aire formado por el movimiento de los discos duros en una computadora, péndulos caóticos e incluso del tipo biométricos \cite{disk,caos,biometrics} pero debido a las inherentes dificultades que ofrece este enfoque, entre las que podemos
mencionar los errores sistemáticos introducidos por el arreglo experimental, la nula reproducibilidad\footnote{En la práctica, frecuentemente se requiere usar muchas veces una misma sucesión de números aleatorios, lo cual implica el  tener que almacenar la sucesión completa con todos los inconvenientes que esto acarrea. Es en este sentido que usamos el término reproducibilidad.} 
de la sucesión obtenida, así como la baja frecuencia en la generación de números aleatorios, han hecho
necesaria  la búsqueda de otras  formas más eficientes para obtener estos números.

\noindent
Desde hace ya algunos años, se utilizan computadoras digitales para implementar programas a los que llamamos
 generadores de números pseudoaleatorios o simplemente generadores, los cuales mediante reglas deterministas
 y  operaciones aritméticas muchas veces sencillas, producen sucesiones de números que se asemejan en un sentido
limitado \cite{knuth,numerical}, a las obtenidas mediante un experimento aleatorio y que se denominan sucesiones
de números \emph{pseudoaleatorios}. 

\noindent
Se conocen muchas implementaciones diferentes para generar números pseudoaleatorios \cite{knuth,numerical}
 que hacen uso de una gran variedad de técnicas y algoritmos que comprenden desde el uso de algoritmos de congruencias
 lineales hasta otros asociados con autómatas celulares, algoritmos de
 criptografía de curvas elípticas, 
etcétera  \cite{automata1,automata2,elipse}.

\noindent
La actual utilización de series muy grandes de  números pseudoaleatorios en muchas aplicaciones, así como
algunos episodios de resultados dudosos, obtenidos debido a la baja calidad de los generadores utilizados
  \cite{numerical,dudas,dudas1,noteasy}, ha fortalecido la necesidad de contar  con mejores y cada vez más eficientes
pruebas de la calidad. El campo de investigación de las pruebas de calidad de generadores de números pseudoaleatorios 
 (y por supuesto, también de su implementación), es tan activo que prácticamente no hay mes en el que no se reporten
 en la literatura científica nuevas pruebas de calidad  que utilizan una gran variedad de  criterios
 y técnicas (teoría de la información, técnicas estadísticas, power spectrum, gambling tests, sistemas físicos, entropía, etc)
 \cite{stat1,informa,stat2,power,gambling,ising,walk,entropia,entropia2}.\\

\noindent
Las pruebas de calidad de los generadores de números pseudoaleatorios se pueden dividir en:
\begin{itemize}
 \item \emph{Pruebas  teóricas}: Se realizan estudiando los algoritmos generadores de números
 pseudoaleatorios mediante el uso de herramientas como la teoría de números. Estos tipos de pruebas son
 útiles por su generalidad y 
están basadas en el estudio de algunas propiedades tales como la longitud del 
periodo de la secuencia y la uniformidad
 del algoritmo. 
\item \emph{Pruebas  empíricas}: Estas pruebas se concentran en 
las sucesiones de números pseudoaleatorios y sus propiedades. Son usadas para
 encontrar correlaciones locales no triviales presentes en las sucesiones de 
números pseudoaleatorios y mostrar aspectos desapercibidos en las pruebas 
teóricas. 
\end{itemize}
En los últimos años se han  explorado pruebas empíricas basadas en procesos
 físicos \cite{ilpo3,ilpo6}, principalmente la caminata aleatoria y el modelo de 
Ising \cite{ising,walk,ilpo3,ilpo6,vattulainen}.\\

\noindent
En este trabajo, se presenta un criterio para evaluar a un generador de
números pseudoaleatorios dado, sugiriendo que su calidad está directamente relacionada con la capacidad
del mismo para producir simulaciones que reproduzcan el comportamiento y propiedades teóricas de un proceso o modelo de referencia. En otras palabras, nuestro enfoque radica en valorar la
calidad de un generador mediante una medida única y global de congruencia entre las propiedades
teóricas del modelo y las producidas por el generador, en lugar de aplicar un conjunto
de pruebas de aleatoriedad aisladas que se basan en desviaciones particulares. Esta idea es aplicada, a manera
de ilustración, para evaluar la calidad de algunos generadores con respecto a su capacidad de simular un proceso
de decaimiento radiactivo. En la siguiente sección haremos un breve
 repaso de la ley del decaimiento radioactivo y de su simulación mediante el Método de Monte Carlo;
 en la sección \ref{generadores} presentamos los generadores que se probarán en el presente trabajo; 
en la Sección \ref{nuestro} explicamos los criterios de nuestro método de prueba de generadores
 de números pseudoaleatorios y en la sección \ref{analisis} ilustramos la aplicación de nuestro método.
  Finalmente, en la sección \ref{results} se hace una discusión de los resultados obtenidos.

\vspace*{-.35cm}
\section{La Ley del Decaimiento Radioactivo}
\vspace*{-.35cm}

\noindent
Por razones de autocontención y con el fin de establecer la notación usada en este trabajo,
 revisemos muy brevemente~\cite{fermi} la ley de decaimiento radioactivo:

\noindent
Considérese al tiempo $t=0$,  una muestra grande con $N(0)$ 
partículas inestables y supongamos que cada partícula tiene una  probabilidad $\lambda  \Delta t$ de
 decaer durante el intervalo de tiempo pequeño $\Delta t$. Aquí $\lambda$ es la constante llamada razón  de decaimiento. Si tenemos un número  $N(t)$  de partículas sin decaer al tiempo t, entonces $\lambda N \Delta t$ partículas  decaerán durante el intervalo de tiempo [$t$, $t$ +$\Delta t$]. 
\noindent
Claramente, esto decrecería el número de partículas que aún no decaen en:
\begin{equation}
 \Delta N = -\lambda N \Delta t
\label{eq:deltan}
\end{equation}
\noindent
Cuando el intervalo de observación en el  tiempo $\Delta t$ tiende a cero, podemos  integrar la ecuación diferencial resultante  obteniendo la ley de decaimiento exponencial:

\begin{equation}
N(t)=N(0) e^{- \lambda t}
\label{eq:decay4}
\end{equation}
\\
\noindent
Las unidades de $\lambda$ son tiempo$^{-1}$.\\

\noindent
Se define la semi-vida como el intervalo de tiempo en el cual la mitad de la muestra inicial habrá decaido: 

\begin{center}
$N(T_{1/2})=N(0)/2$ , es decir:
\end{center} 

\begin{equation}
T_{1/2}=\frac{ln 2}{\lambda}=\frac{0.693}{\lambda}
\label{eq:decay5}
\end{equation}
Finalmente, otra  definición importante a considerar es la de la vida media $\tau$. Esta se define como
 el recíproco de la razón de decaimiento:
\begin{equation}
\tau = \frac{1}{\lambda}
\label{eq:decay6}
\end{equation}

\vspace*{-.35cm}
\subsection{Algoritmo para  Simular la Ley de Decamiento Radiactivo}
\label{algo}
\begin{figure}[!htb]
\resizebox{0.45\textwidth}{!}{%
\includegraphics{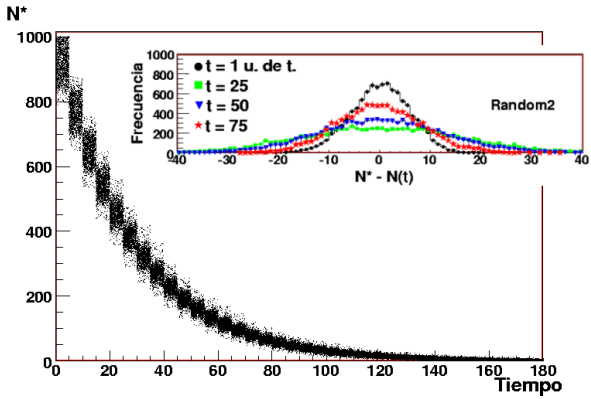}}
\caption{
10 000 simulaciones del decaimiento radioactivo con $\lambda$ = 0.035 , $N_0$ = 1000 partículas y  $\Delta t$ = 0.001; la figura contiene $2\times 10^9$ puntos y se obtuvo usando el generador Random3 descrito en la siguiente sección.
El panel superior muestra la distribución de N*-N(t) para 1,25, 50 y 75 u. de tiempo. Para un buen generador y como consecuencia del Teorema Central del Límite, estos valores se distribuirán gausianamente, lo cual es uno de los criterios de nuestra prueba, como se explicará más adelante en la sección \ref{nuestro}. }
\label{fig:mc}
\end{figure}

\vspace*{-.35cm}
\noindent
El algoritmo utilizado para llevar a cabo la simulación del decaimiento radioactivo es 
el siguiente: 

\begin{enumerate}
\item Se eligen los valores iniciales del decaimiento (cantidad de 
partículas iniciales, valor de $\lambda$, el intervalo de tiempo que se 
simulará $[0,T]$ ) y el tamaño del incremento en el tiempo $\Delta t$. \\

\item Para cada valor discreto del tiempo $ t_i \epsilon  [0,T], i=1.2,..n$; considérense todas
 las $N^{*}(t_i)$ partículas que no han decaido hasta ese instante. Por medio del generador
 de números aleatorios bajo prueba, genérese un número pseudoaleatorio.\\ 

\item Si el número pseudoaleatorio generado en el paso anterior es menor que 
lambda, la partícula decae, y quedarán $N^{*}(t_{i+1}) = N^{*}(t_i)-1$ 
partículas al tiempo $t_{i+1}$. \\

\item Repítase el procedimiento a partir del paso 2 para todas las demás $N^{*}(t_i)-1$
 partículas y para cada $t_i$ en el intervalo especificado. Termínese el procedimiento
 cuando el número de partículas restantes sea menor que 10. (Esto con el fin de evitar
 fluctuaciones estadísticas demasiado grandes).
\end{enumerate}
\noindent
El símbolo * es utilizado de aquí en adelante para indicar que los anteriores
 valores son los valores producidos por la simulación y no por la ley (\ref{eq:decay4}).
 En la figura \ref{fig:mc} se muestra el resultado de realizar esta simulación 10 000 veces,
 con razón de decaimiento $\lambda$ = 0.035 y con un número inicial 
 de partículas de  $N_0$ = 1000.

\vspace*{-.4cm}

\vspace*{-.55cm}
\section{Generadores}
\label{generadores}
\vspace*{-.35cm}

\noindent
En este trabajo, y para ilustrar nuestro método se utilizaron los generadores ran0, ran48 y los generadores  implementados en las clases basadas en C++ TRandom, TRandom2 y TRandom3 y que son
específicos del ambiente ROOT 
\cite{root}, de amplia utilización en física de altas energías. Por simplicidad los llamaremos Random1, Random2 y Random3 y los explicaremos a continuación:\\

\begin{itemize}

\item ran48: Pertenece a la familia de funciones rand48() generadoras de números aleatorios  del lenguaje C estandar y es un generador MLCG (Multiplicative Linear Congruential), es decir,  usa la relación recursiva $r_i$ = $(\alpha+c) $mod $m$, para $i$=$1,2,3...$  Donde  $r_0$, $\alpha$, $m$ y  $c$  son números enteros de 48 bits con valores  $\alpha = 25214903917$, $c = 11$ y  $m=2^{48}$.\\

\item ran0: Es un generador cuya implementación actualizada puede encontrarse en \cite{numerical}  y que fue propuesto por Park \& Miller \cite{Park,Park2} como un modelo de generador con calidad mínima. La implementación usada de este algoritmo genera números pseudoaleatorios con distribución uniforme entre 0 y 1. Se trata de un algoritmo lineal congruencial, donde,  siguiendo la notación de la entrada anterior se hace $\alpha=7^5 = 16807$, $c = 0$ y $m = 2^{31}-1 = 2147483647$.\\

\item Random1: La clase TRandom1, implementa la función Rndm(), que es una 
traducción y actualización  a C++ de  la funcion fortran generadora de 
pseudoaleatorios RNDM, la cual  pertenece a las famosas bibliotecas científicas 
CERNLIB \cite{cernlibs}.  Random1 genera  números pseudoaleatorios 
mediante el método de congruencias lineales, con una distribución uniforme 
en el intervalo [0,1] con un periodo de $10^{8}$. Los parámetros usados por 
esta función son $\alpha=5^{15} = 30517578125$, $c = 0$ y $m = 2^{47} = 
140737488355328$.\\

\item Random2: Implementa un generador de números pseudoaleatorios
 basado en el el método de Tausworthe 
máximamente equidistribuido por L'Ecuyer y usa 3 palabras por estado. Toda la información sobre este generador se 
puede consultar en \cite {ecuyer1,ecuyer2}. El periodo de Random2 es 
$2^{88} = 309485009821345068724781056$.\\

\item  El generador Random3, se  basa en el método ``Mersenne Twister'' en la implementación de M. Matsumoto y  T. Nishimura, \cite{mesentwist}. Random3 es un generador 623-dimensional que produce números
equiprobables en el intervalo [0,1]. Su gran ventaja es su periodo largo
($2^{19937}-1$), aunque su desventaja es su estado interno relativamente
grande de 624 enteros.

\end{itemize}

\noindent 
Los códigos fuente de los tres últimos generadores se pueden consultar en la página de ROOT\cite{root}.

\vspace*{-.35cm}
\subsection{Breve estudio preliminar de la calidad de los  generadores usados en este trabajo}
\vspace*{-.35cm}

\noindent
Con el objetivo de hacer una comparación independiente de nuestra prueba, el cuadro \ref{tab:testgen} muestra los resultados de aplicar 5 sencillas pruebas de aleatoriedad a los generadores usados en nuestro estudio. Estos se realizaron sobre una pequeña muestra de 30 000 eventos para cada generador y consistieron en:

\begin{itemize}

\item Cálculo de $\mu$, el valor esperado promedio. Óptimo valor $\mu$ = 0.5. \\

\item Cálculo de la desviación estandar. Óptimo valor: \\
$\sigma = \sqrt\frac{7}{12} \simeq$ 0.28861.\\

\item Prueba de Anderson-Darling \cite{Darling}. El mejor generador es aquel que minimize al estadístico $A^2$, el cual  nos da la distancia entre la función de distribución empírica y la teórica, en este caso la uniforme en (0,1).\\

\item Estimación de $\pi$ = 3.14159....\\

\item Cálculo del coeficiente de correlación de Pearson $C_p$. Óptimo valor $C_p= 0$.\\

\item Compresión de los archivos de la muestras usando el algoritmo de Lempel~\& Ziv \cite{ziv} y comparando el grado de compresión con una muestra del mismo tamaño de números verdaderamente aleatorios obtenidas de la medición del ruido atmosférico \cite{randomorg}. Óptimo (ideal) valor $0\%$.\\

\end{itemize}

\vspace*{-.4cm}

\begin{table}[!htb]
\begin{center}
\begin{tabular}{|c|c|c|c|}
\hline
Generador & $\mu\pm0.002$ &$\sigma\pm0.0012$ &$A^2$\\
\hline
\hline
Random1 & 0.500& 0.2885 &1.145 \\ 
\hline
Random2 &0.502 & 0.2889 & 0.280\\
\hline
Random3 & 0.496 & 0.2891 & 0.523\\
\hline
drand48 & 0.502 &0.2877 & 0.859 \\
\hline
Ran0 & 0.500 &0.2885 & 1.048 \\
\hline
\hline
\end{tabular}
\end{center}
\end{table}
\vspace*{-1.6cm}
\begin{table}[!htb]
\begin{center}
\begin{tabular}{|c|c|c|c|}
\hline
Generador & $\pi$&$C_p$ & Compresión\\
\hline
\hline
Random1 & 3.1443& 0.0019 & 1\% \\ 
\hline
Random2 & 3.1404 & 0.0003 & 1\% \\
\hline
Random3 & 3.1364&  0.0011 & 1\% \\
\hline
drand48 & 3.1308 & 0.0018 & 1\% \\
\hline
Ran0 & 3.1343 & 0.0089 & 1\% \\
\hline
\hline
\end{tabular}
\caption[]{Algunas sencillas  pruebas de calidad realizadas a una muestra de cada generador de 30 000 números aleatorios uniformemente distribuidos en el intervalo (0,1). Cuadro superior: Test de uniformidad de los generadores. \\
Cuadro inferior: Cálculo de $\pi$ mediante el método de Monte Carlo, Coeficiente de autocorrelación y compresibilidad de la muestras generadas. Los parámetros se explican en el final de la presente sección.}
\label{tab:testgen}
\end{center}
\end{table}


\vspace*{-.35cm}
\section{Criterios de calidad}
\label{nuestro}
\vspace*{-.35cm}

\noindent
Como se indicó en la seccion 1, el criterio básico aquí utilizado para valorar la calidad de un
generador en el contexto de un modelo dado, se relaciona directamente con la capacidad del generador para
reproducirlo y es en ese sentido que presentamos una medida estadística de congruencia, fundamentada en las
propiedades teóricas del proceso bajo simulación. En nuestro caso, consideraremos tres
características: 
\begin{enumerate}
\item Normalidad conjunta de los valores simulados.\\
\item Convergencia de los promedios de los valores simulados a sus promedios teóricos
 (convergencia de primeros momentos).\\
\item Convergencia de las varianzas y covarianzas muestrales a sus valores teóricos
(convergencia de segundos momentos).\\
\end{enumerate}

\noindent
Bajo un proceso de decaimiento exponencial como el que aquí se ha referido, puede verificarse
que el valor esperado y la varianza de $ N^* \left( t_i \right)$, están dados por 
\[
\mu(t_{i})=E\left[ {N^* \left( t_i \right)} \right]=N(0)\exp(-\lambda t_{i}),
\]

\[
\sigma ^2 \left( {t_i } \right) =V\left[ {N^* \left( t_i \right)} \right]= N(0)\lambda \sum\limits_{j = 1}^{t_i } {\exp \left[ { - \lambda \left( {t_i  + j - 1} \right)} \right]},
\]

Además, la covarianza entre las dos variables aleatorias $N^{*}(t_{i})$ y $N^{*}(t_{j})$
se expresa mediante:
\[
\sigma \left( {t_i ,t_j } \right) =
\left( {1 - \lambda } \right)^{t_j  - t_i } \sigma ^2 \left( t_i \right),
\]
donde $t_i  \leq t_j$ y $h_{ij}=t_j-t_i$, $i$, $j = 1,2,....,n$, con $n$ igual a nuestro número de observaciones.\\

\noindent
Para cada uno de los $n$ instantes de tiempo $t_{1},\ldots,t_{n}$ en el intervalo $[0,T]$, denotemos
por ${\bf N^*}=\left[ N^* \left( {t_1 } \right), \ldots ,N^* \left( {t_n } \right) \right]^\prime$
al vector aleatorio que representa los valores del proceso. El vector ${\bf N^*}$ tiene entonces como
valor esperado, al vector
${\bf \mu}  = \left[ \mu \left( {t_1 } \right), \ldots ,\mu \left( {t_n } \right)\right]^\prime$
y por matríz de covarianzas $ {\bf \Gamma}  = \left[ \sigma \left( {t_i ,t_j }
\right) \right].$\\

\noindent
Bajo el supuesto de normalidad conjunta, la cantidad
\[
D = \left[ {{\bf N}^*  - \mu } \right]^\prime  {\bf \Gamma}^{ - 1} \left[ {{\bf N}^*  - \mu } \right],
\]
tiene una distribución $\chi^2$ con $n$ grados de libertad.\\

\noindent
Dadas $m$ realizaciones del proceso, denotemos por ${\bf \bar N^*}$ al vector de promedios
\[
{\bf \bar N^*} = \left[ {\bar N\left( {t_1 } \right), \ldots ,\bar N\left( {t_n } \right)} \right]^\prime,  
\]

\noindent
donde $\bar N^* \left( {t_i } \right) = \frac{1}{m}\sum\limits_{j = 1}^m {N_j ^* \left( {t_i } \right)}$  y
${N_j ^* \left( {t_i } \right)}$ denota el $j-$ésimo valor simulado en el punto $t_{i}$.\\

\noindent
Independientemente del supuesto de normalidad del vector ${\bf N^*}$, por aplicación del Teorema Central
del Límite, el vector $\sqrt m \left( {{\bf \bar N}^*  - {\bf \mu} } \right)$
converge en distribución a la de una variable aleatoria normal $n-$variada con vector de medias ${\bf 0}$ 
y matríz de covarianzas ${\bf \Gamma}$; de ahí que la cantidad
\[
D_{m} = m\left( {{\bf \bar N}^*  - \mu } \right)^\prime
{\bf {\bf \Gamma}}^{ - 1} \left( {{\bf \bar N}^*  - \mu } \right),
\]
tiene como distribución límite, una $\chi^2$ con $n$ grados de libertad.\\
\noindent
$D_{m}$ es una medida estadística del ajuste entre los valores teóricos
de los parámetros de primero y segundo orden derivados del modelo, en este caso
el de decaimiento exponencial,
y sus contrapartes muestrales obtenidos a partir de la simulación. Esta medida
es la que en este trabajo se propone como criterio cuantitativo para la comparación de los
generadores, ya que su valor se incrementa en presencia de desviaciones en los valores teóricos
de los parámetros o bien en presencia de desviaciones de su estructura teórica de covarianzas.
En el contexto de nuestro modelo, el mejor generador será aquél que produzca el menor valor de $D_m$.
Cabe señalar que, con este enfoque, la(s) causa(s) específica(s) de las desviaciones que pudieran
presentarse debidas a deficiencias del generador (como correlaciones, ciclos o cortos períodos, etc) 
nos resultan irrelevantes en tanto que nuestro interés radica en detectar su efecto final, observado
como incongruencia estadística de las simulaciones con respecto a las propiedades del proceso
que se simula.

\vspace*{-.3cm}
\section{Implementación del Método}
\label{analisis}
\vspace*{-.3cm}
\noindent
Ilustremos ahora el procedimiento anteriormente descrito, considerando un proceso
con los siguientes parámetros: $N(0)=1000$ partículas iniciales, constante de decaimiento 
$\lambda$=0.035, $\Delta t$ = 0.0025 y $m = 2000$ simulaciones para cada generador.
Los $n=6$ instantes observados en el intervalo $[0,250]$ fueron $t_1=1$, $t_2=25$, $t_3=50$,
$t_4=75$, $t_5=100$ y $t_6=125$.

\noindent
El cuadro  \ref{tab:resultados} muestra los valores calculados
de $D_m$ para cada uno de los generadores y sus valores de probabilidad (p) asociados, con base en
la distribución $\chi^2$  con 6 grados de libertad.

\begin{table}[!htb]
\begin{center}
\begin{tabular}{|c|c|c|}
\hline
Generador &$D_m$&Probabilidad \\
\hline
\hline
drand48 & 2.914 & 0.820 \\ 
\hline
ran0 & 132.79 & 0.00 \\
\hline
Random1 & 14.500 & 0.025  \\
\hline
Random2 & 2.634  & 0.85 \\
\hline
Random3 &1.223  & 0.976  \\
\hline
\hline
\end{tabular}
\caption{Valores calculados de la estadística $D_m$ para cada uno de los generadores examinados.}
\label{tab:resultados}
\end{center}
\end{table}
\noindent
Para estos valores, el vector de medias y la matríz de covarianzas teóricos
del modelo, son:
\[
\mu  = \left[ {\begin{array}{*{20}c}
   {965.605} & {416.862} & {173.774} & {72.440} & {30.197} & {12.588}  \\
\end{array}} \right]
\]
\[
\Gamma  = \left[ {\begin{array}{*{20}c}
   {33.796} & {14.372} & {5.898} & {2.420} & {0.9933} & {0.4076}  \\
   {} & {247.367} & {101.514} & {41.659} & {17.096} & {7.016}  \\
   {} & {} & {146.104} & {59.958} & {24.605} & {10.097}  \\
   {} & {} & {} & {68.375} & {28.060} & {11.515}  \\
   {} & {} & {} & {} & {29.801} & {12.230}  \\
   {} & {} & {} & {} & {} & {12.649}  \\
\end{array}} \right]
\]

\noindent
El menor valor de $D_m$ corresponde al calculado para el generador Random3, por lo que
concluimos que éste reproduce el modelo de decaimiento radiactivo con mayor precisión.
Aunque el generador Random2 presentó un valor mayor de $D_m$, el valor de probabilidad
asociado nos indica que, aproximadamente un $85\%$ de las veces en que se simule este
proceso bajo las mismas condiciones, podríamos esperar un valor mayor o igual a 2.634,
por lo que el valor obtenido es suficientemente pequeño 
para concluir que la simulación es muy aceptable.
\noindent
Por el contrario, Ran0 y Random1 son los peores generadores según nuestro criterio, el resultado producido por el generador
Random1, nos indica que la probabilidad de obtener un valor tan grande
como $D_m$=14.500 es del orden de 1 en 40, significando con esto que
un valor tan improbable nos lleva necesariamente, a concluir que este
generador no simula satisfactoriamente nuestro proceso. Ran0 es aún peor con un valor de $D_m$ casi un orden de magnitud mas grande que el de Random1 y con probabilidad cero.

\begin{figure}[htb!]
\resizebox{0.45\textwidth}{!}{%
\includegraphics{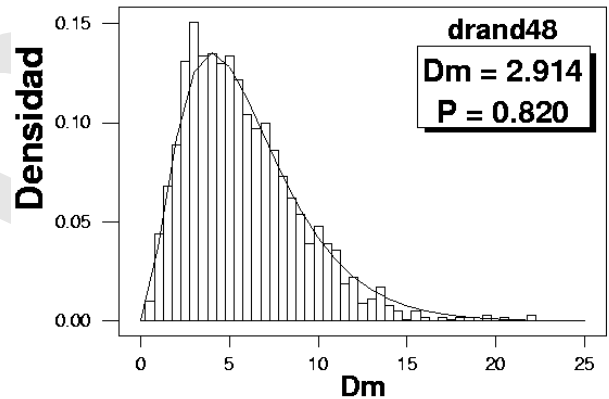}}
\caption{Histograma de los  valores de $D$ para drand48. Se muestra el resultado de  2000 simulaciones. La linea sólida representa la densidad $\chi^2_{(6)}$, que teóricamente tiene $D$. En esta prueba obtenemos los valores $D_m$ =  2.914 con probabilidad $p$ = 0.820.}
\label{fig:rand48}
\end{figure}

\begin{figure}[htb!]
\resizebox{0.45\textwidth}{!}{%
\includegraphics{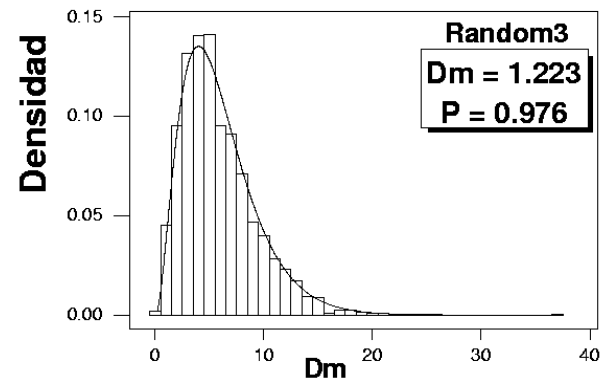}}
\caption{Histograma de $D$ corresponediente al generador Random3, con
$\lambda$ = 0.035 y $n$ =6. , que teóricamente
tiene $D$; para este generador, el ajuste de $\chi^2_{(6)}$ es el mejor de todos, lo cual se ve númericamente de los valores $D_m$ =  1.223 y $p$ = 0.976.}
\label{fig:histod3}
\end{figure}

\noindent
Las figuras  \ref{fig:rand48} a  \ref{fig:histod3} muestran los histogramas de
los valores $D$ bajo las mismas condiciones de simulación para cada uno
de los generadores bajo estudio, recordando que para cada generador, éstos corresponden a 2000 simulaciones del decaimiento radioactivo con $\lambda$ = 0.035, un número de 1000 partículas iniciales e intervalo de tiempo de $\Delta t $= 0.0025 unidades de tiempo.  En ellos se aprecia una congruencia del
valor obtenido de $D_m$ con el ajuste $\chi^{2}_{(6)}$ ilustrado por la
linea continua en cada una de ellas. 

\vspace*{-.45cm}


\begin{figure}[htb!]
\resizebox{0.45\textwidth}{!}{%
\includegraphics{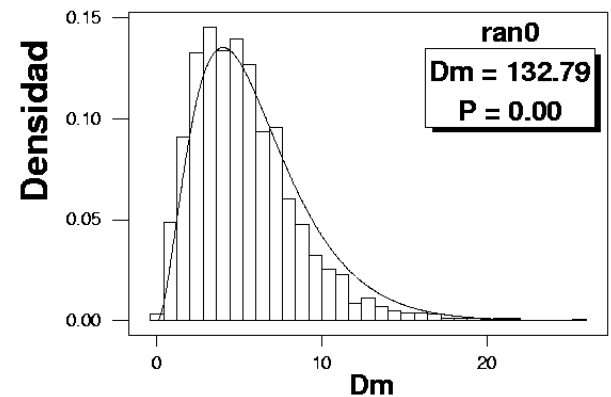}}
\caption{Histograma de $D$ para ran0. La linea sólida representa la densidad $\chi^2_{(6)}$, que teóricamente tiene $D$; note que el ajuste no parece satisfactorio, lo cual es claro de los valores obtenidos de $D_m$ =  132.79 y  probabilidad $p$ = 0.0 en esta prueba.}
\label{fig:ran0}
\end{figure}

\begin{figure}[htb!]
\resizebox{0.45\textwidth}{!}{%
\includegraphics{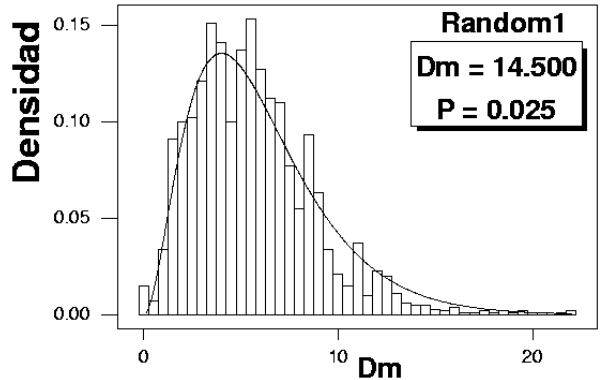}}
\caption{Histograma de $D$ para el generador Random1. Como antes, la linea sólida representa 
la densidad $\chi^2_{(6)}$ teórica de $D$; note que el ajuste no es  satisfactorio, lo cual puede verse de valores obtenidos en esta prueba, los cuales son $D_m$ =  14.500 y una  probabilidad $p$ = 0.025.}
\label{fig:histod1}
\end{figure}

\vspace*{-.45cm}
\begin{figure}[htb]
\resizebox{0.45\textwidth}{!}{%
\includegraphics{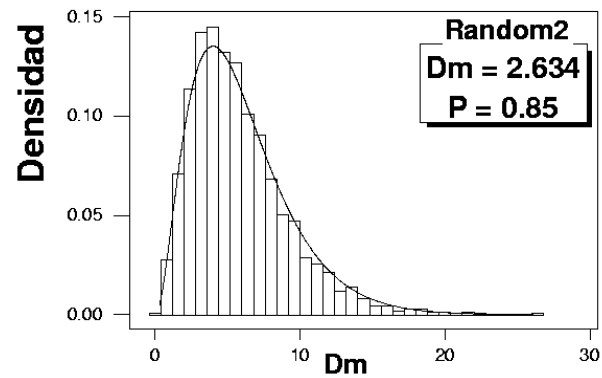}}
\caption{Histograma de  $D$  para el el generador Random2 y su comparación con la linea sólida que representa la densidad de $\chi^2_{(6)}$, que teóricamente tiene $D$; en este caso, el ajuste parece adecuado y obtuvimos los valores de  $D_m$ =  2.634 y $p$ = 0.85}
\label{fig:histod2}
\end{figure}

\noindent
Es importante destacar que, por el Teorema Central
del Límite, la distribución de $D_m$ es poco sensible a desviaciones
de la normalidad de los vectores ${\bf N^{*}}(t)$ y consecuentemente
a la distribución original de $D$; sin embargo, a partir de nuestros
resultados, podría percibirse incorrectamente que
$D_m$ es sensible a desviaciones en la distribución teórica de los
valores $D$, por lo que es necesario aclarar que la congruencia observada entre
nuestra medida de ajuste $D_m$ y las variaciones en la distribución de $D$
con respecto a la distribución $\chi^{2}_{(6)}$, como la que se aprecia
en la figura  \ref{fig:histod1}, son consecuencia de diferencias entre los
valores teóricos de los parámetros y los producidos por la simulaciones,
mismas que se ven reflejadas como irregularidades en el histograma respectivo.
 Es, en este sentido específico, que es posible concluir que $D_m$ mide 
indirectamente esas irregularidades en la distribución de $D$, condensando
el comportamiento general en una sola cantidad.
 
\section{ Conclusiones}
\label{results}
\vspace*{-.4cm}

\noindent
En este trabajo se propone y se fundamenta teóricamente una prueba  empírica de generadores de números pseudoaleatorios basada en el proceso estocástico de decaimiento exponencial. Específicamente hacemos uso de la distribución del número de partículas sobrevivientes y  su normalidad en corcondancia con el teorema central del límite.  Nuestra prueba, tal como debiera esperarse para este tipo de criterios de aleatoriedad, posee la característica de ser sensible tanto a desviaciones distribucionales como a las de los parámetros del modelo usados en la simulación.\\
Se ilustra este método mediante su aplicación a cinco generadores  conocidos: tres generadores congruenciales lineales con distintos parámetros, un Tausworthe y un Mersenne Twister, siendo el mejor de todos según los criterios de nuestra prueba este último.\\
Nuestro método no resulta de difícil  implementación, solo hay que tener cuidado al realizar las simulaciones completas del proceso de decaimiento radioactivo,  con las dificultades computacionales que esto implica, especialmente con lo que respecta al tamaño del intervalo de tiempo elegido (véase \cite{Landau} pg. 53). Por otro lado, es posible, considerando la longitud del periodo de los generadores modernos  usar una muestra relativamente pequeña de sus valores  En este trabajo, para obtener la muestra a analizar con nuestro método, se hicieron 2000  simulaciones del decaimiento radioactivo para cada generador, esto es en total se simularon 50 000 000 eventos y de estos seleccionamos 2000 valores para cada uno de los seis tiempos distintos (1,25,50,75,100 y 125 unidades recíprocas de $\lambda$). Es decir, en nuestra prueba y  para cada generador, de los 50 000 000 eventos simulados un total 12 000 eventos fueron analizados.


\begin{acknowledgement}
{\bf \large Agradecimientos}\\
\noindent
Agradecemos los útiles comentarios hechos a este trabajo por  N. Cruz, S. Jiménez-Castillo y M. Rodríguez-Achach, también agradecemos al Consejo Nacional de Ciencia y Tecnologia (Conacyt) el apoyo brindado  bajo los proyectos No SEP-2003-C02-44598 y 44625.

\end{acknowledgement}


\begin{thebibliography}{}

\bibitem{MC1} N. Metropolis and S. Ulam, J. Amer. Statist. {\bf44}, 335 (1949).

\bibitem{MC2} N. Metropolis, A. Rosenbluth, A. H. Teller and E. Teller. J. Chem. Phys. {\bf 21} 1087, (1953).

\bibitem{stat1} B. Ya, Ryakbo, V.S. Stognienko, Yu.I. Shokin. A new test for randomness and its application to some cryptographic problems. Journal of statistical planning and inference. Elsevier. {\bf123} 365-376 (2004).

\bibitem{crypto1} M. Blum and M. Shub. How to generate cryptographically
 strong sequences of pseudorandom bits. SIAM J. Computing 13 (4) 850-853 (1984).
\bibitem{integracion} J. Bossert, M. Feindt and U. Kerzel. 
Fast integration using quasi-random numbers. Nuclear Instruments and Methods in Physics Research Section A. Volume 559, Issue 1, 1 (2006), 232-236.

\bibitem{eco1} B. Vilenkin. Ecological reading of random numbers. Ecological Modelling, Volume 195, Issues 3-4, 15 (2006), 385-392

\bibitem{identificai}
Loris Nanni and Alessandra Lumini. Human authentication featuring signatures and tokenised random numbers. Neurocomputing, Volume 69, Issues 7-9, (2006), 858-861
\bibitem{identifica2} Dario Maio and Loris Nanni.Multihashing, human authentication featuring biometrics data and tokenized random number: A case study.
Neurocomputing, Volume 69, Issues 1-3, (2005), 242-249

\bibitem{inteligencia}
Navindra Persaud. 
Humans can consciously generate random number sequences: A possible test for artificial intelligence. 
Medical Hypotheses, Volume 65, Issue 2, 2005, 211-214

\bibitem{disk}.D.Davis,R. Ihaka and P. Fenstermacher,  Cryptographic Randomness from air turbulence in disk airs. Proceedings of Crypto 94, Springer Verlac Lecture Notes in Computer Science, No. 839, 1994.

\bibitem{caos}Salih Ergün and Serdar Ozogdcu.  Truly random number generators based on a non-autonomous chaotic oscillator. International Journal of Electronics and Communications, (2006) En prensa.

\bibitem{biometrics} J. Szczepanski et al. Biometric Random Number generators. Computers \& Security, Elsevier, {\bf 23},77-84 (2004).


\bibitem{knuth}Donald E. Knuth. The Art of Computer Programming,  Addison-Wesley Professional. Second edition (1998).

\bibitem{numerical}Numerical Recipes in C: The Art of Scientific Computing
by William H. Press, Brian P. Flannery, Saul A. Teukolsky, William T. Vetterling. Cambridge University Press. Second edition (1992).

\bibitem{automata1} Stephen Wolfram. Random Sequence Generation by Cellular Automata. Advances in Applied Mathematics, {\bf7} 123-169 1986.

\bibitem{automata2} Sheng-Uei Guan, Shu Zhang. Pseudorandom number generation based on controllable cellular automata. FGCS, Elsevier {\bf20} 627-641 (2004).

\bibitem{elipse}Lap-Piu Lee and Kwok-Wo Wong. A Random Number Generator Based on Elliptic Curve Operations. An International Journal of Computers \& Mathematics with applications. Elsevier. {\bf47} 217-226 2004).


\bibitem{dudas} J. R. Heringa, H. W. Blote y A. Compagner. \emph{J. Computat. Phys.}, 10:250, 1983.

\bibitem{dudas1} A. M. Ferrenberg, L.D. Landau and Y.J. Wong. Phys. Rev. Lett. {\bf69} 3382 (1992).

\bibitem{noteasy} P. Hellekalek. Good random number generators are (not so) easy to find. \emph{Mathematics and Computers in Simulation}, 46:485-505, 1998.


\bibitem{informa} B.Ya. Ryabko and V.A. Monarev.
Using information theory approach to randomness testing
Journal of Statistical Planning and Inference, Volume 133, Issue 1, 1 July 2005, Pages 95-110


\bibitem{stat2} Makoto Matsumoto and Takuji Nishimura. Sum-discrepancy test on pseudorandom number generators. Mathematics and computers in simulation. Elsevier. {\bf62} 431-442 (2003).

\bibitem{power} Nezih C. Geclinli, Murat A. Apohan. Power spectrum tests of random numbers. Signal Processing. Elsevier. {\bf81} (2001) 1389-1405.

\bibitem{gambling} Stefan Wegenkittl. Gambling test for pseudorandom number generators. Mathematics and Computers in Simulation. Elsevier. {\bf55} 281-288 (2001). 

\bibitem{ising}I. Stauffer. Ising Model as test for simple random number generators. International Journal of Modern Physics C. {\bf5} 807-808 (1999).



\bibitem{walk} Mihyun Kang. Efficieny test of Pseudorandom number generators using random walks. Elsevier. Journal of Computational and Applied Mathematics, 174 1 2005.


\bibitem{entropia} P. L'Ecuyer.  Entropy Tests for Random Number Generators. \emph{GERAD report}, 1996.

\bibitem{entropia2} Andrew L. Rukhin. Aproximate Entropy for Testing Randomness. J. Appl. Probab. {\bf37} 1 88-100 (2000).




\bibitem{ilpo3} I. Vattulainen, T. Ala-Nissila, and K. Kankaala, "Physical Models as Tests of Randomness", Physical Review E vol. 52, 3205 (1995).

\bibitem{ilpo6} I. Vattulainen, K. Kankaala, and T. Ala-Nissila. "Physical Tests for Random Numbers in Simulations", Physical Review Letters vol. 73, 2513 (1994).

\bibitem{vattulainen} I. Vattulainen,  \emph{New tests of random numbers for simulations in physical systems.} Tesis, Universidad de Helsinki, Helsinki, Finlandia, 1994.

\bibitem{fermi} E. Fermi,Nuclear Physics. University of Chicago Press. Chicago (1949) USA. 

\bibitem{Park} S. Park and K. Miller. Random number generators: good ones are hard to find. Comm. ACM 31:1192--1201, 1988.

\bibitem{Park2}S. Park and K. Miller. Comm. ACM 36 No. 7, 105-110, 1993.


\bibitem{cernlibs}
CERN Program Library. CERNLIB. Short Writeups. Application Software and Databases. Computing and Network Division. Edition  june 1996. CERN Geneva, Switzerland.

\bibitem{ecuyer1} P. L'Ecuyer. Mathematics of Computation, 65, 213 (1996).

\bibitem{ecuyer2} P. L'Ecuyer.  Mathematics of Computation, 68, 225 (1999). Véase:\\www.iro.umontreal.ca/$\sim$lecuyer/myftp/papers/tausme.ps. 

\bibitem{mesentwist} M. Matsumoto y T. Nishimura. Mersenne Twister: A 623--dimensionally equidistributed uniform pseudorandom number generator. \emph{ACM Transactions on Modeling and Computer Simulations: Special Issue on Uniform Random Number Generators}. Vol. 8, No. 1, January 1998, pp 3-30. Más información:\\
http://www.math.keio.ac.jp/$\sim$matumoto/emt.html.


\bibitem{root} Rene Brun and Fons Rademakers, ROOT - An Object Oriented Data 
Analysis Framework, Proceedings AIHENP'96 Workshop, Lausanne, Sep. 1996, Nucl. 
 Inst. Meth. in Phys. Res. A 389 (1997) 81-86. Ver también http://root.cern.ch/.
\bibitem{Darling} R.B. D'Agostino and M.A. Stephens, Goodness of fit Techniques Marcel Dekker, New York, (1986).

\bibitem{ziv} J. Ziv and A. Lempel, A Universal Algorithm for Sequential Data Compression.  IEEE Transactions on Information Theory, Vol. 23, pp. 337--342, 1977.

\bibitem{randomorg} La muestra fue descargada del el sitio www.random.org.

\bibitem{Landau} D. P. Landau and K. Binder. A guide to Monte Carlo Simulations in Statistical Physics. Cambridge University Press (2000).

\end{thebibliography}
\end{document}